# Fast emitting nanocomposites for high-resolution ToF-PET imaging based on multicomponent scintillators.

*Matteo Orfano, Fiammetta Pagano, Ilaria Mattei, Francesca Cova, Valeria Secchi, Silvia Bracco, Edith Rogers, Luca Barbieri, Roberto Lorenzi, Gregory Bizarri, Etiennette Auffray, Angelo Monguzzi\**

M. Orfano, F. Cova, V. Secchi, L. Barbieri, A. Monguzzi
Dipartimento di Scienza dei Materiali, Università degli Studi Milano-Bicocca, via R. Cozzi 55, 20125 Milano, Italy

E-mail: angelo.monguzzi@unimib.it

F. Pagano, E. Auffray
CERN, Geneva, Switzerland.

E. Rogers, G. Bizarri
Cranfield University, School of Aerospace, Transport & Manufacturing, MK43 0AL, UK

I. Mattei
INFN Sezione di Milano, via G. Celoria 16, 20133, Milan, Italy

**Abstract.** Time-of-Flight Positron Emission Tomography is a medical imaging technique, based on the detection of two back-to-back γ-photons generated from radiotracers injected in the body. Its limit is the ability of employed scintillation detectors to discriminate in time the arrival of γ-pairs, i.e. the coincidence time resolution (*CTR*). A *CTR* < 50 ps that would enable fast imaging with ultralow radiotracer dose. Monolithic materials do not have simultaneously the required high light output and fast emission characteristics, thus the concept of scintillating heterostructure is proposed, where the device is made of a dense scintillator coupled to a fast-emitting light material. Here we present a composite polymeric scintillator, whose density has been increased upon addition of hafnium oxide nanoparticles. This enhanced by +300% its scintillation yield, surpassing commercial plastic scintillators. The nanocomposite is coupled to bismuth germanate oxide (BGO) realizing a multilayer scintillator. We observed the energy sharing between its components, which activate the nanocomposite fast emission enabling a net *CTR* improvement of 25% with respect to monolithic BGO. These results demonstrate that a controlled loading with dense nanomaterials is an excellent strategy to enhance the performance of polymeric scintillators for their use in advanced radiation detection and imaging technologies.

# 1. Introduction

Time-of-Flight Positron Emission Tomography (ToF-PET) is a powerful imaging technique that provide reliable information in a variety of medical fields, from oncology to neurology.[1] Briefly, a radiotracer labelled with a β- radioactive isotope (a positron emitter) is injected into the patient and then it is attracted to areas of increased metabolic activity, such as tumours. The isotope undergoes radioactive decay, releasing a positron that annihilates with a surrounding electron producing two back-to-back γ photons with 511 keV energy. As showed in **Figure 1a**, these γ-rays are sensed by scintillating detectors placed around the patient's body. The recorded signals are therefore used to reconstruct the image of the radio-labelled tissue. One of the limiting factors of this technique is the detectors ability to discriminate in time the arrival of the back-to-back γ-photons pairs. The uncertainty on the position $\Delta x$ of the annihilation event along the back-to-back line of response is given by

$$\Delta x = \frac{c \Delta t}{2}, \qquad \text{Eq. 1}$$

where $c$ is the speed of light in vacuum and $\Delta t = t_2 - t_1$ is the uncertainty on the different detection times of the two γ photons, i.e. the coincidence time resolution ($CTR$, Fig.1b).[2] The $CTR$ is therefore a crucial figure of merit for a ToF-PET detector. Using monolithic scintillators, it can be estimated from several parameters intrinsic of the material as

$$CTR = 3.33 \left(\frac{\tau_{rise}\tau_{eff}}{N}\right)^{0.5} = 3.33 \left(\frac{\tau_{rise}\tau_{eff}}{\beta \chi [\Phi_{scint} E]}\right)^{0.5}, \qquad \text{Eq. 2}$$

where $\tau_{rise} = (1.57\tau_{rise}^0 - 1.15\tau_{opt})$ is the global scintillation pulse rise time constant given by the intrinsic scintillation rise time $\tau_{rise}^0$ of the material plus the average optical photon transit time in the device to the photodetector $\tau_{opt}$.[3] The effective emission time $\tau_{eff} = (k_{dtc})^{-1}$ is the reciprocal of the detection rate of scintillation photons $k_{dtc}$ (Supporting Information, section 1). $N$ represent the absolute light output, i.e. the number of photons collected by the photodetector. The light output can be expressed as function of several parameters specific of any scintillator, including the amount of energy deposited in the scintillator $E$, the scintillator light yield $\Phi_{scint}$ that indicate the number of photons generated per MeV of deposited energy, the light outcoupling efficiency $\beta$ that set the fraction of scintillation photons that reach the photodetector, and the photodetector quantum efficiency $\chi$.

As suggested by Eq. 1, the $CTR$ should be as low as possible to improve the instrument sensitivity and the image signal-to-noise ratio at low radiotracers concentration, [4] thus potentially allowing a broad use of safe, low-dose and fast ToF-PET screening protocols and opening the way also to yet unexplored application in neurology and paediatrics. State-of-the-art commercial machines works with $CTR$ = 210 ps, [5] so much effort is then put in the development of new scintillators aiming at a $CTR$ of 10 ps that would allow fast imaging with acquisition times shorter with respect to the standard clinical using very low limited radiotracer dose.[6] Unfortunately, none of the existing monolithic scintillators shows simultaneously the high $\Phi_{scint}$ and the sub-nanosecond $\tau_{eff}$ necessary to reach the 10 ps $CTR$. High $\Phi_{scint}$ are typical of slow emitting heavy crystals,[7] while organic and plastic scintillators have fast emission also in the sub-nanosecond range[8] but a low density inadequate to stop the 511 keV γ-rays in the small ToF-PET detectors architecture.

In order to overcome this obstacle, the concept of scintillating heterostructure has been proposed.[9] [10] Here the scintillator is fabricated by coupling heavy crystalline scintillators, such as the bismuth germanate oxide (BGO), to fast-emitting plastic scintillators realizing a multicomponent device (Fig.1c). The dense component stops γ-rays by means of photoelectric effect, thus enabling the selection of the correct back-to-back 511 keV events. The fast emission is activated directly by charges recombination and, importantly, by the energy deposited by diffusing recoil photoelectrons electrons generated in the dense component that reaches the plastic component. This energy sharing mechanism is the key to possibly obtain simultaneously the energy resolution and the sub - 200 ps $CTR$ that is required to further improve the quality and speed of the ToF-PET image reconstruction. Unfortunately, organic materials, polymers and plastics have typical densities of $\rho \approx 1$ g cm$^{-3}$, thus the presence of the fast emitter decreases the global density of the scintillating heterostructure and therefore reduces its effective stopping power at 511 keV.[11]

To mitigate this effect, we developed a composite polymeric scintillator whose density was artificially increased by loading it with high density nanoparticles. In particular, we employed polystyrene (PS) as host matrix for the 1,4-bis(5-phenyl-2-oxazolyl)benzene (POPOP) scintillating dye.[12]-[13] As high density additive, we employed hafnium oxide (HfO$_2$, $\rho$ = 9.68 g cm$^{-3}$) nanoparticles.[14] In an optimized composition, we achieved a 300% improvement of $\Phi_{scint}$ surpassing the performance of several commercial plastic scintillators. The nanocomposite has been therefore employed as fast emitting component in the manufacturing of a multilayer heterostructure coupled to crystalline BGO. The prototype device properties have been investigated by means of photoluminescence and scintillation spectroscopy experiments. We observe the occurrence of energy sharing and a net improvement of the timing performance with respect to the reference monolithic BGO detector. The investigation of the scintillation process kinetics revealed an unexpected deceleration of the fast nanocomposite emission rate due to a still unclear interaction of the guest dye with the host matrix, which partially compensate the achieved gain in the material scintillation yield. The obtained results demonstrate that a controlled loading with dense nanoparticles is an excellent strategy to enhance the $\Phi_{scint}$ of plastic scintillators based on fast emitting dyes, as well as the light output of scintillating heterostructures by partially keeping the stopping power required. On the other hand, the spectroscopic analysis performed clearly suggests that a more detailed knowledge and control of the excited states interactions during scintillation in the sub-nanosecond time scale are still to be achieved, in order to preserve the dyes emission properties to maximize the device timing performance.

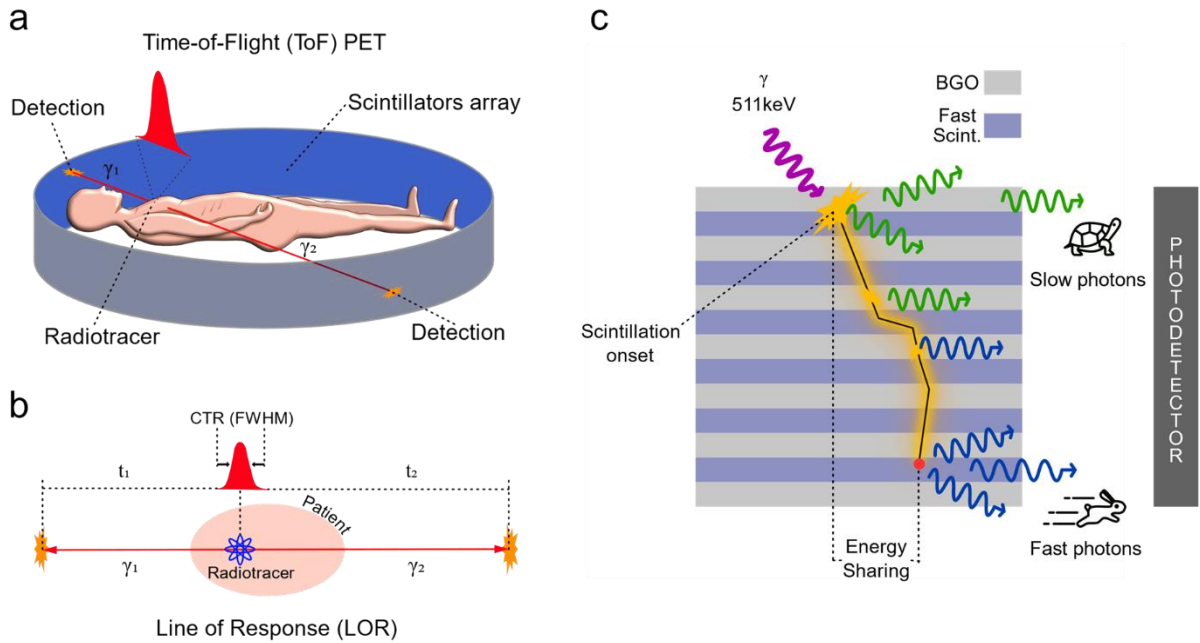

**Figure 1.** a) Sketch of the basic principles of ToF-PET imaging technique. b) Elements use for the determination of the ToF-PET scanner coincidence time resolution (CTR), which is defined as the average uncertainty $\Delta t$ of the time difference $t_2 - t_1$ in the detection of the back-to back γ-rays (γ$_1$ and γ$_2$) along a selected line of response (LOR) in the detector ring. c) Sketch of the multilayer scintillators built with a series of heterostructure obtained by coupling the heavy scintillating bismuth germanate oxide (BGO) layers and a fast plastic scintillator layer. The fast emission of this latter is activated by exploiting the energy sharing form the photoelectric secondary recoil electrons produce in the BGO upon interaction with the γ-rays.

## 2. Optical and scintillation properties of fast emitting nanocomposites.

**Figure 2a** sketches out the nanocomposite structure. The PS matrix has been obtained by a thermal-assisted polymerization of styrene monomers, in which POPOP ($10^{-2}$M) and HfO$_2$ nanoparticles have been dispersed. The POPOP is a scintillating conjugated chromophore with a photoluminescence quantum yield $\Phi_{pl}$= 0.93 and a photoluminescence lifetime $\tau_{pl}$ = 1.4 ns (Fig.S1, Supporting Table S1).[15] As shown in Fig.2b, different processes lead to visible emission upon interaction of the polymeric matrix with the ionizing radiation.[16] We have both the direct recombination of diffusing charges on the POPOP molecules and the host-guest non radiative energy transfer from the PS matrix to POPOP molecules (Fig.S2). In the proposed composition, 98% of the PS excitons transfer their energy to POPOP molecule with $\Phi'_{ET} = 1$ and an energy transfer rate $k_{ET} \geq 5.6$ GHz = (180 ps)$^{-1}$. The remaining 2% of PS excitons transfer their energy with yield $\Phi''_{ET} = 0.82$ (Supporting Information section 3, Fig.S3). This is ascribed to a slightly non-homogenous dispersion of POPOP molecules. The total energy transfer yield become therefore $\Phi_{FS} = 0.98\Phi'_{ET} + 0.02\Phi''_{ET} \approx 1$, observed with both optical or X-rays excitation. The full-organic PS:POPOP sample shows a $\Phi_{scint}$ = 3800 ph MeV$^{-1}$ (Methods, Fig.S4). To densify the polymeric scintillator and improve $\Phi_{scint}$, we loaded it with HfO$_2$ nanoparticles (Methods).[17] The HfO$_2$ nanoparticles were synthesized by hydrothermal route (Methods, Supplementary Figs. S4-S6).[17] As showed in Fig.2c, the nanoparticles have an oval shape, with size of 85 nm and 35 nm for the larger and shorter diameter, respectively. The nanoparticles show a negligible weak blue emission,[18] thus their primary role remains the enhancement of the interaction with ionizing radiation thanks to the high atomic number of hafnium (Z=72).

A series of scintillating nanocomposites PS:POPOP-xx%wt have been fabricated loading different amounts of nanoparticles (xx%wt) up to 3% in weight. As shown in Fig.2c, the scintillation intensity and $\Phi_{scint}$ progressively increase with the loading level reaching a maximum for the PS:POPOP-2.5%wt sample (inset), with an excellent $\Phi_{scint}$ of 9500±475 ph MeV$^{-1}$ that surpasses the 8600 ph MeV$^{-1}$ yield of the reference plastic scintillator EJ-286$^{TM}$ (Fig. S7). This composition increases the material density from 1.02 g cm$^{-3}$ to 1.5 g cm$^{-3}$. $^{13}$C solid-state NMR, performed on polystyrene nanocomposites, is a method of choice to highlight the formation of polymer and the absence of residual monomer. Only the characteristic signals of the polymer at 40.7 ppm, 45.9 ppm, 128.5 ppm and 146.3 ppm for the CH, CH$_2$, phenyl-CH and phenyl-C for polystyrene are observed in the $^{13}$C CP MAS NMR spectrum (Fig.S8). Further evidence of complete polymerization is provided by differential scanning calorimetric analysis. In fact, the detection of glass transition in the DSC runs at about 97 °C for PS-POPOP-2.5% and 96 °C for PS-POPOP nanocomposites, respectively, suggested that the same degree of polymerization has been reached for both polymeric materials (Fig. S9). Thermogravimetric analyses, performed under oxidative conditions, enabled us to evaluate the thermal stability of polymeric nanocomposites. The degradation process starts at about 350°C for PS-POPOP-2.5% nanocomposite, 25°C much higher than the decomposition temperature of PS-POPOP, indicating the $HfO_2$ nanoparticles enhance the thermal stability of the polystyrene and dye (Fig. S10).

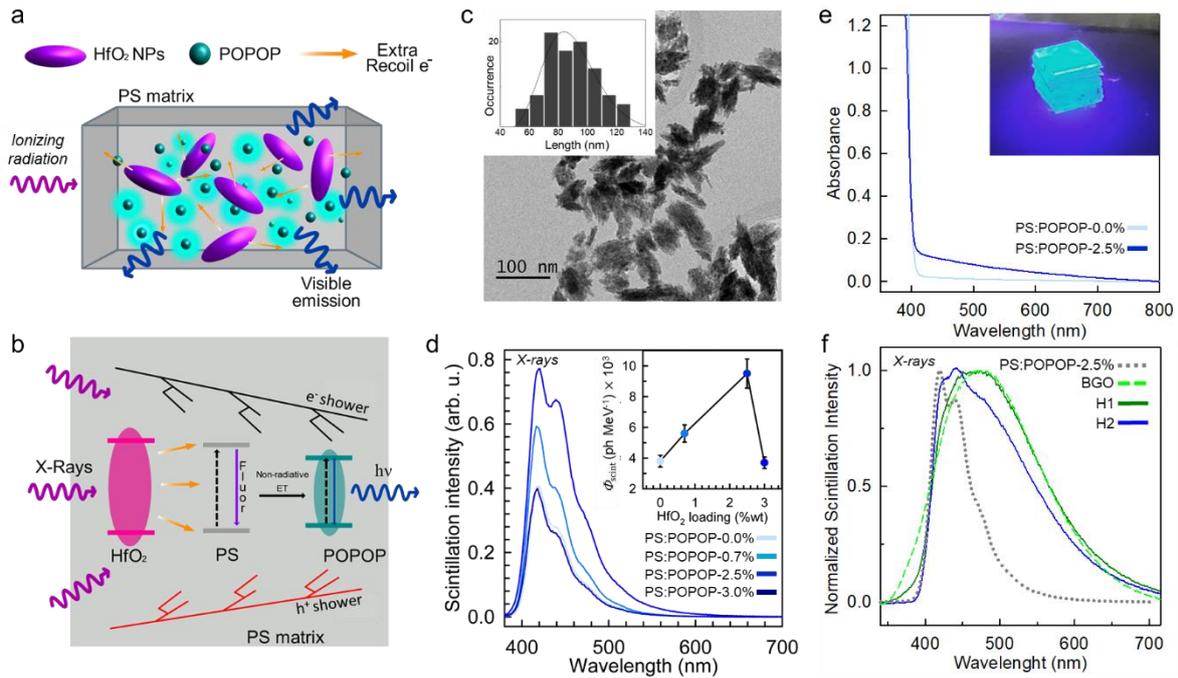

**Figure 2.** a) Sketch of the multicomponent polystyrene (PS) scintillating nanocomposite loaded with the POPOP dye ($10^{-2}$ M) and $HfO_2$ nanoparticles (NPs). b) Outline of the photophysics involved in the nanocomposite scintillation process. c) Transmission electron microscopy (TEM) image of NPs. The inset depicts the NPs log axis size distribution obtained by TEM image analysis. d) Scintillation emission spectrum of the nanocomposite PS:POPOP-x% as a function of the NPs loading in weight (%wt) under exposition to soft X-rays. The inset shows the scintillation light yield ($\Phi_{scint}$) measured as function of the loading level. e) Normal incidence optical absorption of 100 μm nanocomposite film with and without the NPs loading (2.5%wt). The inset is digital picture of a 3×3×3 $mm^3$ multilayer scintillator made of 100 μm BGO and 100 μm nanocomposite alternating layers. f) Scintillation emission spectrum of BGO, PS:POPOP-2.5% and the corresponding single heterostructure under soft X-rays. The heterostructure is excited in two configurations, by hitting first the BGO (H1) and hitting first the nanocomposite (H2).

By further increasing the nanoparticles amount both the emission intensity and $\Phi_{scint}$ decrease significantly, thus suggesting that the 2.5%wt concentration is the maximum acceptable value before the nanoparticles starts to become competitive recombination centres.[16, 19] The addition of nanoparticles only slightly modifies the light transport properties of the polymeric scintillator. From optical absorption experiments (Fig.2d) we estimate a scattering background as high as 0.15 of apparent absorbance for a 100 μm film with a 2.5%wt loading (Fig.2e), with no significant effect on the material light output.[20] Both the scintillation spectrum (Fig.2c) and kinetics (*vide infra*, Fig.5b) do not change in presence of nanoparticles, thus demonstrating that, when used with the correct amount, they act as perfectly passive scintillation sensitizers.[19a, 21] We ascribe the observed sensitization effect to an enhanced conversion yield of the energy deposited by the ionizing radiation into emissive states upon interaction with the dense particles, for example by enhancing the local production of secondary events. None clear explanation for this effect has been found in the literature, [22] and a dedicated study is still ongoing. It is worth noting that in this case $\Phi_{scint}$ is higher than for several commercial plastics, but also it matches the scintillation yield of BGO crystals ($\Phi_{scint}^{BGO}$ = 10000 ph MeV$^{-1}$)[23] that will be used as dense component to fabricate the multilayer scintillating heterostructure. This will avoid any intrinsic unbalance in the generation of the scintillation photons, therefore making Eq. 2 valid also in the multilayer device.

As the fundamental building block of the multilayer device, a single heterostructure unit has been fabricated by coupling a 100 μm film of PS:POPOP-2.5%wt nanocomposite with a 100 μm thick BGO layer (Methods). Figures 2f shows the emission properties of the single heterostructure compared to the ones of the individual components. According to the low penetrability of the radiation in the dense component by exciting with soft X-rays the BGO layer the fast emitter cannot be activated nor directly neither by energy sharing, thus the scintillation emission spectrum matches the one of bulk BGO crystal (H1 spectrum). On the other hand, X-rays cannot be fully stopped by the plastic, so by hitting as first the fast emitter a fraction of X-rays can excite the BGO layer. Therefore, in this case the heterostructure emission spectrum is given by the convolution of PS:POPOP-2.5% and BGO (H2 spectrum).

**3. Observation of energy sharing in the multilayer scintillator.**
The multilayer scintillating heterostructure has been realized by duplicating the heterostructure to reach a final size of 3×3×3 mm$^3$ (Methods). **Figure 3a** shows the pulse height spectra measured for the multilayer device compared to the one of a bulk BGO crystal and a bulk nanocomposite of the same size, under exposure to 511 keV γ-rays (Methods). The plot shows how the heterostructure spectrum has an intermediate behaviour compared to the one of BGO, where a clear photoelectric peak can be observed, and the one of the nanocomposite, where only the broad energy distribution due to the Compton effect appears. The pulse height spectrum of the multilayer still shows a photoelectric peak, but the ratio between photopeak and Compton events is lower due to the decreased stopping power of the heterostructure compared to bulk BGO (Supporting Table S3). Moreover, the photoelectric peak of the heterostructure is shifted toward lower values compared to bulk BGO because of the worse light transport and collection in the multilayer vs. the monolithic device.[24] The correlation between the amplitude and time-integrated intensity of the scintillation energy signal of the different pulse shapes recorded was used to distinguish between events depositing energy in BGO, in the nanocomposite, or when the energy is shared.[24] The time-integrated pulse height spectra for the reference samples is depicted in Fig.3b. As expected considering the nominal different emission lifetimes, 1.4 ns vs. 300 ns for nanocomposite and BGO, respectively (*vide infra*), their response in the time domain is clearly distinguishable by a pulse shape discrimination analysis (Methods). The multilayer scintillator shows a complex behaviour, with a spectrum that contains three regions of interest, as depicted in Fig.3c. The majority of events occur in the left-hand region marked with the green dotted line, where incoming photons interact with the heavy BGO and the recoil photoelectron is fully contained there (Fig.3b, top). The second region is in the lower part, marked with the blue dotted line, which point out the events that occur solely in the nanocomposite (Fig.3b, bottom). The triangular area between the two straight lines is attributed to the scintillation shared events described above, thus indicating that the global recorded output signal is not just the sum of two independent components but that there is an effective interaction between the different materials in this preferred architecture (Figs.S11, S12). According to Monte Carlo simulations of the radiation matter interaction in the system (Methods) the average fraction of energy loss in the nanocomposite with respect to the total energy loss in the whole detector results as high as $\eta = 0.35$. This value if 20% higher than the case of a multilayer scintillator built using a pure polymeric fast scintillator, thus highlighting the role of $HfO_2$ nanoparticles in enhancing the stopping power of the polymeric component (Table S3). A more refined data analysis allows to distinguish the events useful for imaging reconstruction, i.e. the fast light pulses generated in the nanocomposite by recombination of the recoil electrons produced solely by photoelectric effect in the BGO. As shown in Fig.3d, the time resolved pulse height spectrum of the multilayer can be shown using a smart coordinates change from the [time-integrated signal] vs [amplitude] to the [energy in BGO] vs. [energy in nanocomposite] system. The straight lines mark the region where we find the events where the energy deposited in BGO lies between 440 keV and 665 keV (24% of the total, Table S3) and at least 50 keV are shared in the nanocomposite, which enables the best performance for the image reconstruction.[24-25]

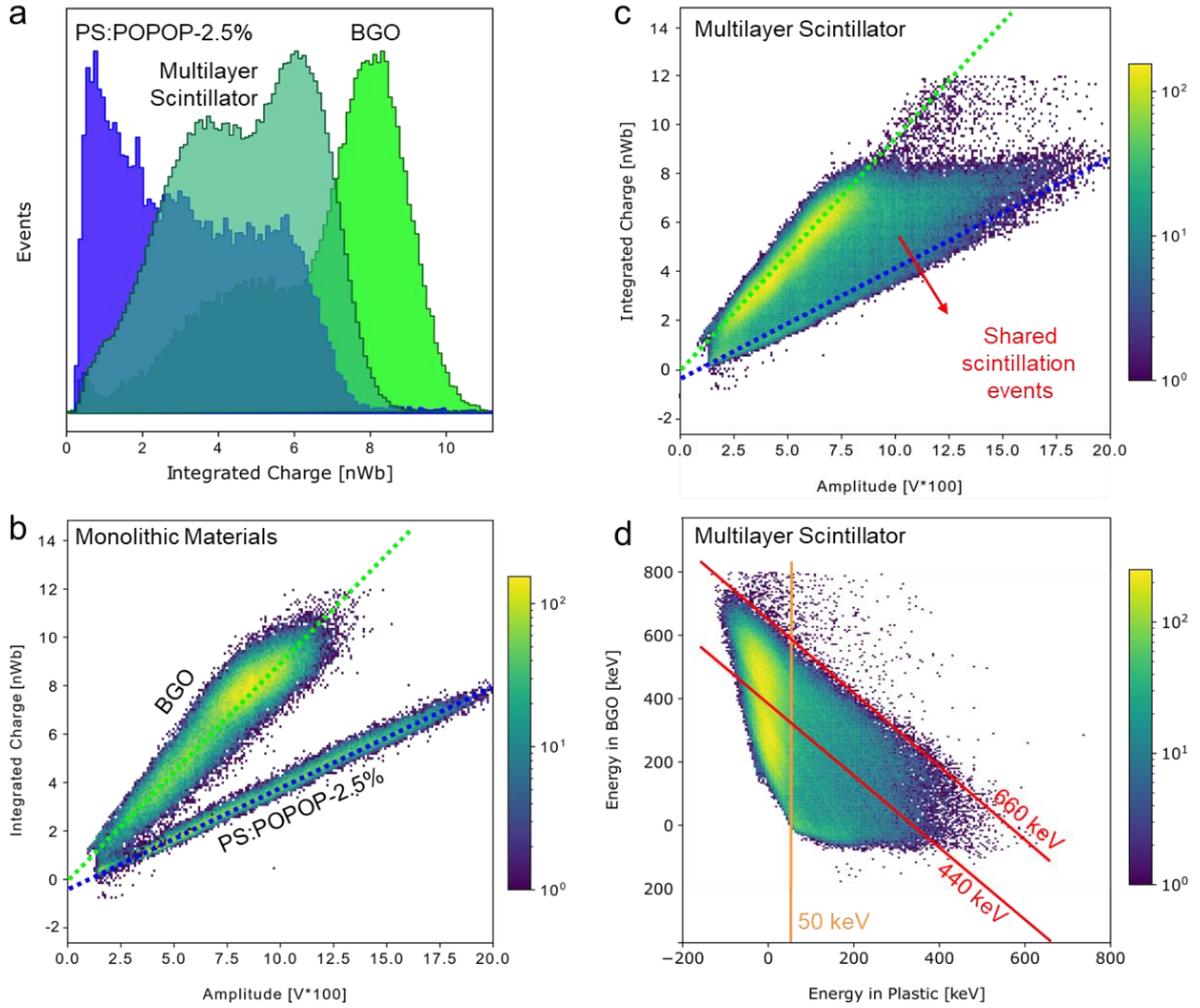

**Figure 3.** a) Pulse height spectrum of BGO, PS:POPOP-2.5% nanocomposite and the multilayer scintillator upon exposure to 511 keV γ-rays readout with a SiPM. b) Time integrated pulse-height spectra for bulk BGO overlapped to that one of monolithic PS:POPOP-2.5% composite. Dotted lines are a guide for the eye. c) Time integrated pulse-height spectrum of 3x3x3 mm³ multilayer scintillator made alternating 100 mm layers of BGO and PS:POPOP-2.5%. Green and blue dotted straight lines are a guide for the eye to highlight the events occurring solely in BGO and nanocomposite, respectively, as like as in panel b. The shaded area between the straight lines marks the shared scintillation events. d) [energy in BGO] vs. [energy in nanocomposite] spectrum of the multilayer scintillator. Straight lines are the energy threshold values employed to select the shared fast scintillation events that originated from BGO photoelectric recoil electrons diffusing in the nanocomposite. As photoelectric events we consider those with total reconstructed deposited energy between 440 to 660 keV. We consider as shared photopeak only the events where at least 50 keV of energy is deposited in the nanocomposite by recoil electrons.

**4. Time resolution performance and scintillation kinetics.**

The time resolution of the multilayer scintillator has been tested by means of ultrafast pulsed X-rays excitation and under exposure to 511 keV γ-rays. It is worth noting that the global distribution in time of scintillating pulses depends on several factors, such as the scintillation rise time and the effect of the device size and optical quality on the light propagation and photons paths, which set the pulses arrival time on the photodetector. Therefore, the reported data are specific of the considered sample series, where device geometry and instrumental setup have been fixed. As preliminary test, the detector time response has been measured under picoseconds X-ray pulsed excitation for the series of devices analysed (Methods).[26] As showed in **Figure 4a**, the multilayer device scintillation is significantly faster than the BGO and similar to that one of the bulk nanocomposite. The same is observed for the distribution in time of scintillations pulses rise time on the photodetector (Fig.4b). The intermediate behavior between the monolithic BGO and nanocomposite demonstrate again a synergistic global response of the two components in the device. **Figure 4c** reports the results of the $CTR$ measurements. In detail, the $CTR$ is obtained as the full width half maximum value of the statistical distribution of the recorded time differences in the detection of 511 keV back-to-back γ-rays emitted by ²²Na atoms (Methods).[27] The experiment has been performed on the whole scintillators series, and again the multilayer scintillator shows an intermediate behaviour with a good $CTR$ of 180 ps that lies between the value of 244 ps for BGO and 141 ps for the bulk nanocomposite, respectively. Thanks to the energy sharing mechanism that activates the fast emission of the nanocomposite component, we successfully observe a clear improvement +26% of the device $CTR$ with respect to bulk BGO, thus achieving a value close to the limit observed in the monolithic nanocomposite reference and very similar to what has been recently obtained using high density fast scintillators such as 2D-perovskite crystals.[9b]

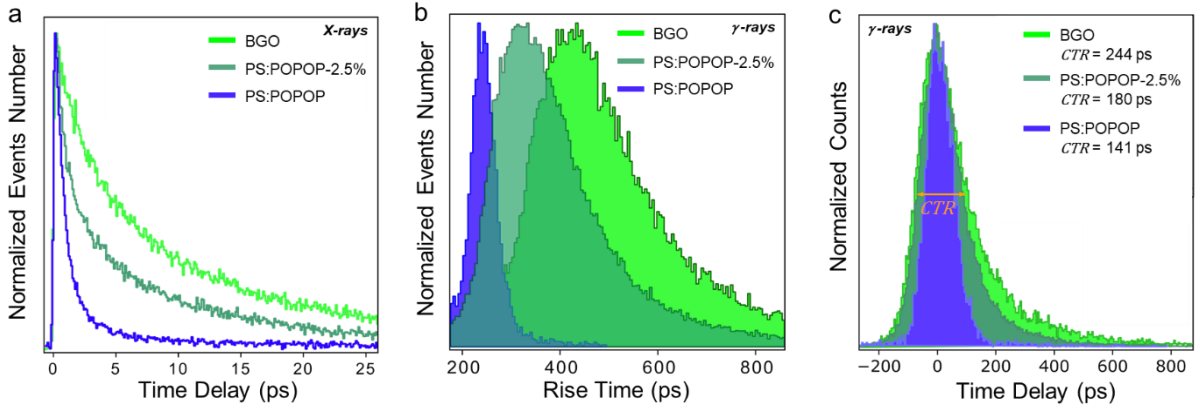

**Figure 4.** a, Detector time resolution measured under soft x-rays on in 3×3×3 mm³ bulk BGO, monolithic PS:POPOP-2.5% nanocomposite and on the multilayer scintillator made made alternating 100 mm layers of BGO and PS:POPOP-2.5% cubic heterostructure of size 3×3× mm³. b, Distribution of the scintillation signal rise times on the photodetector measured for the BGO, PS:POPOP-2.5% composite and the scintillating heterostructure under 511 keV γ-rays exposure. c, Experimental $CTR$ of bulk BGO, monolithic PS:POPOP-2.5% and scintillating heterostructure measured under 511 keV γ-rays exposure.

In order to shed light on which are the $CTR$ limiting factors among the ones included in Eq.2, we investigated the intrinsic scintillation properties of its components. The multicomponent device global time response results actually as the linear combination of the single components' responses, weighted by the fraction of energy deposited in each one of them.[28] More in detail, we can reconstruct the global time response $CTR^†$ by using

$$CTR^† = 3.33 \left(\frac{\tau_{rise}\bar{\tau}_{eff}}{\beta\chi[\Phi_{scint}E]}\right)^{0.5} = 3.33 \sum_i A_i \left(\frac{\tau_{rise}\tau_{eff}^i}{\beta\chi[\Phi_{scint}E]}\right)^{0.5}, \qquad \text{Eq. 3}$$

where the device global scintillation decay time $\bar{\tau}_{eff}$ is given as the linear combination of all the possible $\tau_{eff}^i$ for each $i$-shared event. The $\tau_{eff}^i$ values depends on the effective scintillation decay time in the composite ($\tau_{eff}^{nanoc}$) and in the BGO ($\tau_{eff}^{BGO}$) by

$$\tau_{eff}^i = \left[\frac{\varepsilon_i}{\tau_{eff}^{nanoc}} + \frac{(1-\varepsilon_i)}{\tau_{eff}^{BGO}}\right]^{-1}. \qquad \text{Eq. 4}$$

Their combination in Eq. 3 is weighted by the distribution of probability $A_i$ that the $i$-shared event deposits in the nanocomposite a fraction $\varepsilon_i$ of the BGO photoelectric recoil electrons energy ($\varepsilon_i$ = energy deposited in the nanocomposite $E_i$ / 511 k eV).

By directly measuring the properties of the individual materials, it is therefore possible to model the device $CTR^\dagger$ with good accuracy through Eqs. 3 and 4. Figure 5a shows the scintillation pulse rise kinetics measured in a BGO layer vs. the nanocomposite film. While for BGO the rise time is instantaneous, i.e. below the experimental resolution and therefore negligible, the nanocomposite shows an intrinsic rise time as large as $\tau_{rise}^0$=129 ps (Methods). This value is in agreement with the estimated rate of the PS-to-POPOP energy transfer, and it should be used to calculate the global $\tau_{rise}$ as described above. Considering a device of size 3×3×3 mm³ and scattering coefficient $\alpha_s$ = 0.232 cm⁻¹ at 420 nm (Fig. 2f), we have a $\tau_{opt}$ = 85 ps given by the distribution of photons' optical paths before detection.[20] In this case, the combination of $\tau_{rise}^0$ and $\tau_{opt}$ results a device $\tau_{rise}$ as high as 298 ps. The effective scintillation decay time $\bar{\tau}_{eff}$ is calculated by combining scintillation experiments with Montecarlo simulations to evaluate the $A_i$ and $\varepsilon_i$ coefficients values, which are reported in the Supporting Table S4 (Methods and Supporting Information, section 7). Notably, the simulation results indicate that also the energy sharing effect is enhanced by the presence of nanoparticles with respect to a pure plastic scintillator, with a global increment of the fast scintillation emission by 20%. Figure 5b shows the scintillation pulses recorded for the BGO layer and for the nanocomposite film. Both materials show a multi-exponential decay of the emission intensity with time, from which we obtain a $\tau_{eff}^{BGO}$ = 97.9 ns and $\tau_{eff}^{nanoc}$ = 3.69 ns for the BGO and nanocomposite, respectively.[26, 29]

The theoretical best time response $CTR_{th}^\dagger$ can be now estimated as a function of the measured parameters, considering the certified $\chi$ = 0.5 at 420 nm for the photodetector and the ideal $\beta$ = 0.65 for the selected experiment configuration (Methods).[20] The ideal $CTR_{th}^\dagger$ results as low as 94 ps. Considering that the only free parameter in Eq. 3 is the optical coupling factor $\beta$, we ascribe the difference with respect to the experimental result to a not ideal transport and extraction of photon within the scintillator to the photodetector. As suggested by the scintillation experiments discussed above (Fig.3a), the observed $CTR^{\dagger}$ value of 180 ps corresponds indeed to an effective low light outcoupling yield of $\beta_{eff} = 0.2$. This indicates that a large part of the scintillation photons is lost because of the device poor light transport properties, most probably due to the presence of many non-ideal interfaces between the layers composing the prototype.

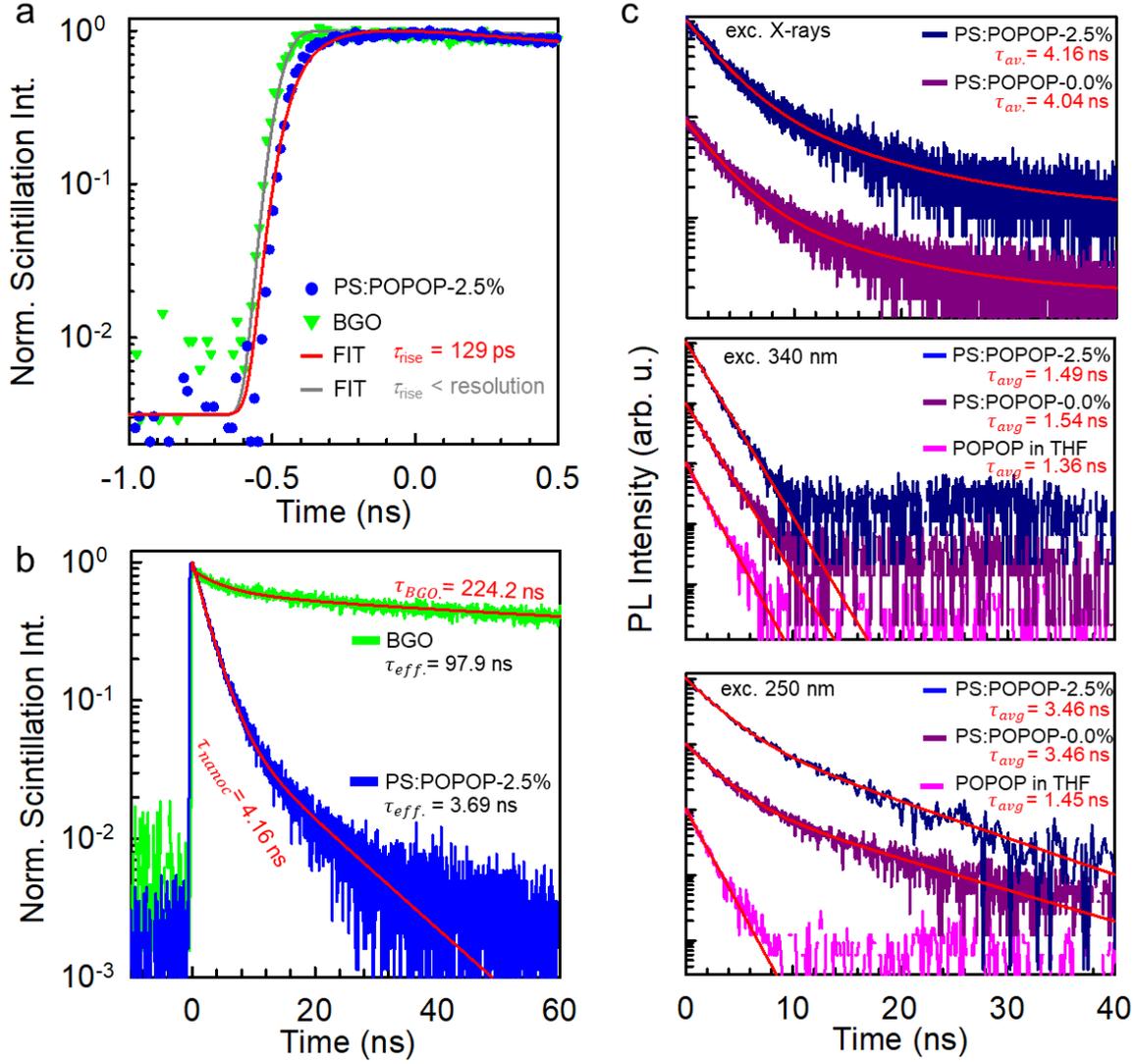

**Figure 5.** a, Rise time of the scintillation pulse measured for BGO and PS:POPOP-2.5% composite under ultrafast excitation with soft X-rays. Solid lines are the fit of the experimental data convoluted with the instrumental time response. b, Scintillation pulse decay in time under excitation with soft X-rays in BGO and PS:POPOP-2.5% nanocomposite. Solid lines are the fit with multi-exponential decay functions with characteristic average decay time $\tau_{avg}$. The effective decay time reported $\tau_{eff}$ from the multi-exponential fit as described in the text. c. Top panel reports the luminescence decay in time at 420 nm recorded in PS:POPOP and PS:POPOP-2.5% composites under pulsed X-ray excitation. Middle and bottom panels show the luminescence decay in time recorded at 420 nm in PS:POPOP and PS:POPOP-2.5% vs. POPOP in tetrahydrofuran solution under pulsed laser excitation at 340 nm and 250 nm, respectively. The solid lines are the fit with single- or multi-exponential decay functions with characteristic average decay time $\tau_{avg}$.

Despite a non-optimized light transport/extraction, the time response of the multilayer scintillator is significantly improved with respect to the bulk and slow BGO scintillator, thanks to the highly efficient and fast scintillation achieved in the nanocomposite. However, we notice that the nanocomposite scintillation decay time is surprisingly longer than the POPOP intrinsic emission lifetime (Fig.5b). Indeed, while the BGO scintillation average decay time is 224.2 ns in agreement with literature values for both scintillation and photoluminescence,[26, 29] the nanocomposite scintillation decays in 4.1 ns, i.e. more than twice than the POPOP photoluminescence. This means that, according to Eq. 3, the potential device time response is intrinsically worsened by a significant factor ∼ 1.5. In order to understand the origin of this discrepancy, we investigated the nanocomposite scintillation and photoluminescence properties in parallel. Figure 5c shows the nanocomposite scintillation pulse decay measured in presence and absence of nanoparticles. No differences can be observed, thus demonstrating that nanoparticles do not affect the emission kinetics acting only as passive radiosensitizers. Interestingly, upon selective direct optical excitation of POPOP at 340 nm (Fig. S1), even if embedded in PS its emission decays in 1.4 ns, as like as a single molecule in solution and independently from the presence of nanoparticles (Fig.5c, mid panel). This also exclude potential effects of self-absorption on the excited states recombination kinetics. Conversely, the POPOP emission generated by energy transfer from the optically excited PS matrix at 250 nm, shows a clear increment in the emission lifetime up to 3.46 ns. This finding suggests therefore that the scintillation lifetime slowdown could depend on an interplay between the excited PS matrix and embedded POPOP molecules, affecting their radiative recombination rate. This picture is further supported by the fact that the energy transfer from PS-to-POPOP is fast enough to not affect the dye excited state recombination kinetics (Fig.S13). Moreover, trivial effects such as the reduction of intramolecular vibrational quenching upon embedding in polystyrene can be excluded, because the POPOP has a photoluminescence quantum yield close to unity indicating the absence of this dispersion mechanism.[30] We speculate that the POPOP scintillation lifetime increment could be possibly due to local polarization of the excited matrix around the excited dyes,[30-31] or to the involvement of dark metastable states, which can act as intermediate energy reservoir, such as the polystyrene lowest energy triplet state $T_1$ at 388 nm that is strictly resonant with the dye $S_0$-$S_1$ transition at 380 nm eV (Fig. S1) or other types of defect and energetic traps in the host [30, 32]. A dedicated study is ongoing to elucidate this point, because it the observed slowdown of the emission can be dramatically detrimental in the search of the sub-nanosecond timing performance.

## 5. Conclusion

To summarize, we successfully fabricated a new fast emitting and high efficiency scintillating polymeric composite based on a polystyrene matrix loaded with scintillating conjugated dyes and high density nanoparticles to enhanced its stopping power. The loading level has been tuned to maximize the nanoparticle radiosensitization effect without introducing competitive deactivation mechanism. This allowed to increase the material scintillation yield by a factor 300%, surpassing several commercial plastic scintillators. The nanocomposite has been employed to fabricate a fast emitting multilayer scintillator as a prototype of the optical part of the detector that could be used in ToF-PET scanners for high-contrast fast imaging. The device has been realized by alternating nanocomposite films and crystalline BGO sheets as dense material to efficiently stop the ionizing radiation and activate the energy sharing mechanisms to trigger the fast nanocomposite emission. Upon exposure to γ-rays at 511 keV, we observe a synergistic response of the two components in the multilayer devices. Thanks to the effective activation of the nanocomposite fast emission upon energy sharing, we obtained a final coincidence time resolution of 180 ps, significantly better than for monolithic BGO (244 ps). The investigation of the scintillation process in the multicomponent device pointed out that its time response is limited by two main factors. The first one is the poor light transport and outcoupling of the scintillation photons. This can be straightforwardly improved by developing an accurate industrial manufacturing of the prototype device, still handmade. The second factor is more fundamental and deserve great attention. The analysis of the scintillation and luminescence processes in the nanocomposite suggest an unexpected interplay between the host matrix and the embedded dyes upon excitation at high energy. This mechanism, the origin of which is still to be clarified, result in a significant slowdown of the dyes luminescence decay rate by more than a factor of three, thus heavily affecting the time response of the device. This is a critical point that should be understood and managed to fully preserve the excellent fast emission properties of organic scintillators that can be exploited to achieve the highly desired time resolution below 50 ps for ToF-PET scanners.

## 6. Methods

*Preparation and structural characterization of HfO$_2$ nanoparticles.* The HfO$_2$ nanoparticles were synthesized by hydrothermal route.[17] The standard experimental procedure is described as follows. The hafnium hydroxide chloride (Hf(OH)$_2$Cl$_2$) solution was firstly prepared by dissolving 0.160 g of HfCl$_4$ in 10.0 mL of deionized water. NaOH aqueous solution (3.0 M, 10.0 mL) was added dropwise to the solution above, causing the reaction with Hf(OH)$_2$Cl$_2$ to form hafnium hydroxide (Hf(OH)$_4$). After that, the solution was transferred into a 100 mL Teflon-lined autoclave and the sealed autoclave was heated to 120 °C and maintained for 72 hours. The products were purified by centrifugation for three cycles with alcohol and deionized water alternately after the autoclave was cooled down. Finally, the precipitate was dried at 50 °C for 24 hours. The structure and composition of HfO$_2$ nanoparticles were studied by means of powder X-ray diffraction (PXRD) structure refinement, Raman spectroscopy and transmission electron microscopy (TEM). Details and data are reported in the Supplementary Information file, section 2. XRD and Raman spectra show a well-defined series of diffraction peaks and Raman bands in good agreement with the values of monoclinic hafnium dioxide.[33],[34]

*Preparation of polystyrene-based nanocomposites.* Styrene monomer was purchased from Sigma-Aldrich (CAS no. 100-42-5) under the form of a liquid and colourless monomer. After the removal of stabilizer, the 1,4-Bis (5-phenyl-2-oxazolyl) benzene (POPOP, Sigma Aldrich, CAS no. 1806-34-4, MW=364.40 g/mol) dye was dissolved with the addition of the as-prepared hafnium oxide nanoparticles. Then, the monomer polymerization followed a thermal pathway by using Azobisisobutyronitrile (AIBN), a free radical initiator, the VAZO$^{TM}$ 64 (Chemours$^{TM}$). The final composition was obtained as follows: 3.64 mg of POPOP and 1 mg of VAZO$^{TM}$ 64 were dissolved in 1 ml of styrene through ultrasonic stirring. Then 25 mg of hafnium oxide nanoparticles were added to the solution, and then dispersed through stirring. The as-prepared solution was placed in a temperature-controlled oil bath at 80°C for 1 day. For the first 8 hours, the mixture was mechanically stirred every hour to avoid nanoparticles sedimentation. At the end of the process, we obtained a polystyrene-based plastic scintillator with a 10-2 M dye concentration and 2.5% wt hafnium oxide nanoparticles loading.

*Preparation of nanocomposites films.* 100 μm polystyrene films loaded with 10$^{-2}$ M 1,4-Bis (5-phenyl-2-oxazolyl) benzene and 2.5% wt of hafnium oxide nanoparticles were prepared as follows: 1g of composite prepared as discussed in the section above was dissolved in 1 ml of DCM (Sigma Aldrich, CAS no. 75-09-2). Once the solution reached a syrup-like viscosity it was deposited with the help of a doctor blade using a 5 mils blade. The film was then left air drying until the complete solvent evaporation.

*Assembling of scintillating heterostructures.* The 3 mm × 3 mm × 0.1 mm BGO sheets, purchased from EPIC Crystal LTD, and 0.1 mm nanocomposite films were assembled together by hand using un-polymerized styrene monomer as a glue, realising an ultra-thin layer acting as adhesive for the composite films. Once the latter sheet has been placed in position, the whole heterostructure was left under chemical hood for the monomer to self-polymerize. For more accurate measurements and to maximize the device quality, the multilayer scintillator has been fabricated by alternating BGO and nanocomposite 0.1 mm layers, with no glue, in a custom made sample holder, specifically a 3×3×3 mm Teflon cube with one face open, which internal surface has been covered by ESR (Vikuiti) form 3M as reflecting material.

*Photoluminescence studies.* Time-resolved photoluminescence experiments in the nanosecond time scale have been performed by using as excitation source a pulsed laser LED at 340 nm (3.65 eV, EP-LED 340 Edinburgh Instruments, pulse width 120 ps) and a pulsed laser LED at 250 nm (4.95 eV, EP-LED 250 Edinburgh Instruments, pulse width 77 ps) coupled to FLS980 Edinburgh setup in Time-Correlated Single Photon Counting (TCSPC) acquisition mode. Quartz Suprasil cuvettes with 1 cm of optical path has been used for all the experiments listed above to study dyes solution. The nanocomposites were excited with a pulsed laser at 340 nm avoiding the excitation of the polystyrenic matrix, coupled to FLS980 Edinburgh setup in Time-Correlated Single Photon Counting (TCSPC) acquisition mode.

*Radioluminescence studies.* Steady state RL measurements were carried out at room temperature using a homemade apparatus featuring, as a detection system, a liquid nitrogen-cooled, back-illuminated, and UV-enhanced charge coupled device (CCD) Jobin-Yvon Symphony II, combined with a monochromator Jobin-Yvon Triax 180 equipped with a 100 lines/mm grating. All spectra are corrected for the spectral response of the detection system. RL excitation was obtained by unfiltered X-ray irradiation through a Be window, using a Philips 2274 X-ray tube with tungsten target operated at 20 kV. At this operating voltage, a continuous X-ray spectrum is produced by a Bremsstrahlung mechanism superimposed to the L and M transition lines of tungsten, due to the impact of electrons generated through thermionic effect and accelerated onto a tungsten target. The dose rate was 5 mGy/ s, evaluated with an ionization chamber in air.

*X-rays decay.* The scintillation emission rate was studied in time correlated single photon counting (TCSPC) mode under pulsed X-ray excitation. For this purpose, X-ray Tube (XRT) N5084 of Hamamatsu was used. The X-rays energy spectrum is a bremsstrahlung continuous spectrum extending up to 40 keV (as the operating voltage is 40kV) with an additional pronounced peak around 9keV due to Tungsten L-characteristic X-ray photons. As photodetector. As photodetector, a hybrid photomultiplier tube (HPM 100-07 from Becker&Hickl) optimized for TCSPC measurement was used. The samples were measured in anti-reflection positioning. [26] The scintillation emission rates were fitted with the convolution between the overall impulse response function (IRF) of the system and the intrinsic scintillation rate. The former was obtained by the analytical convolution between the measured IRF of the laser together with HPM and the IRF of the X-ray tube resulting around 160 ps FWHM. The latter was modelled with a sum of bi-exponential functions.

*DTR measurements.* The excitation branch of the system is the same as for scintillation decay measurements, the difference is in the detector. Here a SiPM (from Hamamatsu, 53 V breakdown voltage, 61 V bias voltage) is used, its signal is split in two in order to optimise independently the energy and time information. The first one is processed by an analog amplifier, while the latter by a custom made high-frequency amplifier (two cascade radio frequency BGA616 amplifiers).[27] The outputs are then digitized at the oscilloscope (Lecroy, WaveRunner 8104, 20 Gs/s sample rate, 1 GHz bandwidth), where all information required for the analysis is measured and extracted directly from the waveforms. The samples were wrapped with one layer of Teflon positioning them on the SiPM without applying any glue or grease for the optical coupling.[26]

*CTR measurements.* The CTR under 511keV was measured with a standard setup as previously described.[27] In brief, a 22Na radioactive source emits two back-to-back 511 keV gamma photons which are detected by two detectors in coincidence. On one side a reference crystal (LSO:Ce:Ca0.4% 2x2x3 mm$^3$, 61ps CTR FWHM), on the other side the sample under investigation (BGO 3x3x3 mm$^3$, POPOP:HfO$_2$ 3x3x3 mm$^3$, heterostructure BGO+POPOP:HfO 3x3x3 mm$^3$). Both samples are coupled to a SiPM through Meltmount glue (1.58 refractive index). For the samples under investigation was a Hamamatsu SiPM (53 V breakdown voltage, 61 V bias voltage) was used. The readout electronics was the same used also for DTR measurements. The output signals feed an oscilloscope (LeCroy DDA735Zi oscilloscope with 3.5 GHz bandwidth and a sampling rate of 40 Gs s$^{-1}$), where all information required for the analysis is measured and extracted directly from the waveforms.

*Monte Carlo Simulations.* In order to evaluate the expected energy sharing between the slow BGO crystal and fast scintillator component, a Monte Carlo simulation of the experimental setup has been performed by means of the FLUKA code. [35] The geometry of the multilayer scintillating heterostructure with a final size of 3×3×3 mm$^3$, as well as the two components materials (BGO and POPOP:HfO$_2$), in terms of atomic weights and density, have been fully reproduced. A 511 keV photon isotropic source has been simulated, at a distance of 0.5 cm from the detector surface. Dedicated user-routines have been developed in order to analyse the simulation output on an event-by-event basis, providing the information of the energy deposits in the detector. No optical simulation has been performed at this stage.

*Scintillation pulse simulations.* Scintillation pulse simulations and fitting have been performed considering deconvolution of Instrument Response Function (IRF) with rise and decay times. The temporal profile of scintillation can be described as:

$$I(t) = (f * g)(t) := \int_0^t f(\tau) g(t-\tau) d\tau$$

Where f(t) and g(t) represent the IRF and a combination of rise and decay functions, respectively. The temporal profile of IRF has been assumed to be Gaussian in shape:

$$f(t) = \frac{e^{-\frac{(t-t_0)^2}{2w_t^2}}}{\sqrt{2\pi} w_t}$$

Where $t_0$ and $w_t$ are the temporal position of the peak maximum and the standard deviation of the IRF. The latter includes the duration of both the x-ray excitation and the detector response. In our apparatus, the FWHM of the IRF is equal to 110 ps corresponding to a deviation standard $w_t \approx$ FWHM/2.35=47 ps.

The time decay of the sample has been modelled as the product of a rising exponential and a decaying exponential:

$$g(t) = \left(1 - e^{-\frac{t}{\tau_R}}\right) e^{-\frac{t}{\tau_D}}$$

Considering the definitions of *f(t)* and *g(t)*, the convoluted signal can be represented as:

$$I(t) = \int_0^t f(\tau) g(t-\tau) d\tau = \int_0^t \frac{e^{-\frac{(\tau-t_0)^2}{2w_t^2}}}{\sqrt{2\pi} w_t} \left(1 - e^{-\frac{(t-\tau)}{\tau_R}}\right) e^{-\frac{(t-\tau)}{\tau_D}} d\tau$$

Which has the following analytical solution:

$$I(t) = c + \frac{1}{2} e^{\frac{w_t^2 + 2\tau_D(t_0 - t)}{2\tau_D^2}} \left( \text{Erf}\left[\frac{w_t^2 + t_0 \tau_D}{\sqrt{2} w_t \tau_D}\right] - \text{Erf}\left[\frac{w_t^2 + (t_0 - t)\tau_D}{\sqrt{2} w_t \tau_D}\right] \right) +$$

$$+ \frac{1}{2} e^{\frac{(\tau_D + \tau_R)[2(t_0 - t)\tau_D \tau_R + w_t^2(\tau_D + \tau_R)]}{2\tau_D^2 \tau_R^2}} \left( \text{Erf}\left[\frac{(t_0 - t)\tau_D \tau_R + w_t^2(\tau_D + \tau_R)}{\sqrt{2} w_t \tau_D \tau_R}\right] - \text{Erf}\left[\frac{t_0 \tau_D \tau_R + w_t^2(\tau_D + \tau_R)}{\sqrt{2} w_t \tau_D \tau_R}\right] \right)$$

Where $t_0$ and $w_t$ are known and define the IRF shape. The parameters used to fit normalized scintillation pulse data are c, $\tau_R$, and $\tau_D$ which represent the constant dark background, and the rise and the decay constants, respectively. Simulations, reported in Fig. S13, have been calculated assuming *c=0, t₀=0.5 ns* and *w$_t$ = 47 ps*, keeping fix $\tau_D$ at 1.5 ns, and varying $\tau_R$ from 10 ps to 10 ns.

## Supporting Information

Supporting Information is available from the Wiley Online Library or from the author.

## Acknowledgements

This work was carried out in the framework of Crystal clear Collaboration. We acknowledge support from the European Community through the grant no. 899293, HORIZON 2020 - SPARTE FET OPEN. Financial support from the Italian Ministry of University (MUR) through grant no. PRIN 2020—SHERPA no. H45F2100343000. CERN knowledge transfer for medical applications budget. Cranfield University, acknowledges that part of this research was funded by the UK Engineering and Physical Sciences Research Council (EPSRC) grant EP/S013652/1.

We thank Professor Anna Vedda, for valuable discussions.

**TOC**

We realized a new composite fast emitting polymeric scintillator loaded with high density nanoparticles to enhance its scintillation performance. Thanks to the achieved 300% scintillation yield increment, the nanocomposite has been successfully employed to fabricate a prototype multilayer scintillating heterostructure for high resolution ToF-PET imaging with a < 200 ps time response.

*Matteo Orfano, Fiammetta Pagano, Ilaria Mattei, Francesca Cova, Valeria Secchi, Edith Rogers, Luca Barbieri, Gregory Bizarri, Roberto Lorenzi, Etiennette Auffray, Angelo Monguzzi\**

**Fast emitting nanocomposites for high-resolution ToF-PET imaging based on multicomponent scintillators.**

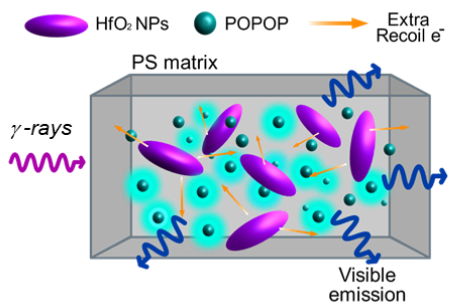
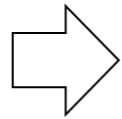
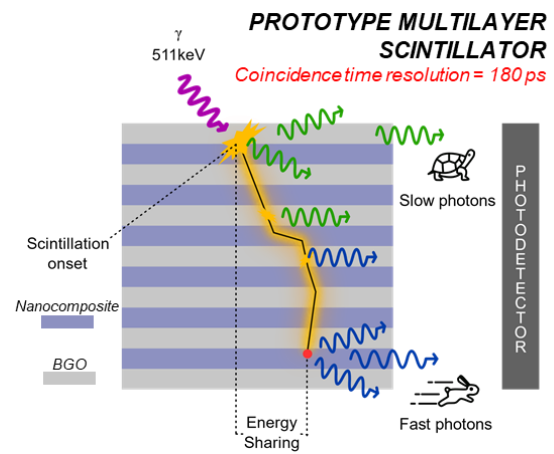

Supporting Information

**Fast emitting nanocomposites for high-resolution ToF-PET imaging based on multicomponent scintillators.**

*Matteo Orfano, Fiammetta Pagano, Ilaria Mattei, Francesca Cova, Valeria Secchi, Edith Rogers, Luca Barbieri, Gregory Bizarri, Roberto Lorenzi, Etiennette Auffray, Angelo Monguzzi\**

**INDEX**

## 1. Time resolved data analysis.

The time resolved and scintillation measurements in the main text show a complex behaviour that is typically fitted using multi-exponential functions as

$$I_{PL}(t) \propto \sum_i B_i e^{-(t/\tau_i)}.$$ Eq. S1

The average emission characteristic lifetime has been calculated as

$$\tau_{avg.} = \sum_i \frac{B_i \tau_i}{B_i}.$$ Eq. S2

The effective emission decay time is calculated had the harmonic weighted average of the weighted emission decay component using

$$\tau_{eff.} = (k_{dtc})^{-1} = (\sum_i A_i k_i)^{-1} = \left(\sum_i \frac{A_i}{\tau_i}\right)^{-1},$$ Eq. S3

where is $A_i$ is the relative weight of each decay rate component $k_i = (\tau_i)^{-1}$.

**Table S2.** Fit parameters for the PL and scintillation emission intensity decay curves with time.

| **Exc. 340 nm** | A₁ | $\tau_1$ [ns] | A₂ | $\tau_2$ [ns] | A₃ | $\tau_3$ [ns] | $\tau_{avg}$ [ns] |
|---|---|---|---|---|---|---|---|
| POPOP in THF | 1.0422 | 1.3571 | - | - | - | - | 1.3571 |
| PS:POPOP | 1.0433 | 1.5368 | - | - | - | - | 1.5368 |
| PS:POPOP:HfO₂ | 1.0603 | 1.4901 | - | - | - | - | 1.4901 |

| **Exc. 250 nm** | A₁ | $\tau_1$ [ns] | A₂ | $\tau_2$ [ns] | A₃ | $\tau_3$ [ns] | $\tau_{avg}$ [ns] |
|---|---|---|---|---|---|---|---|
| POPOP in THF | 1.0322 | 1.4507 | - | - | - | - | 1.4507 |
| PS:POPOP | 0.8540 | 2.3994 | 0.1619 | 9.0582 | - | - | 3.4610 |
| PS:POPOP:HfO₂ | 0.8439 | 2.4467 | 0.1768 | 7.7431 | - | - | 3.3641 |

| **Soft X-rays** | A₁ | $\tau_1$ [ns] | A₂ | $\tau_2$ [ns] | A₃ | $\tau_3$ [ns] | $\tau_{avg}$ [ns] |
|---|---|---|---|---|---|---|---|
| PS:POPOP | 0.7482 | 2.8660 | 0.1370 | 10.4629 | - | - | 4.0419 |
| PS:POPOP:HfO₂ | 0.7596 | 2.9476 | 0.1378 | 10.8272 | - | - | 4.1578 |

## 2. Optical and luminescence properties of POPOP and polystyrene (PS).

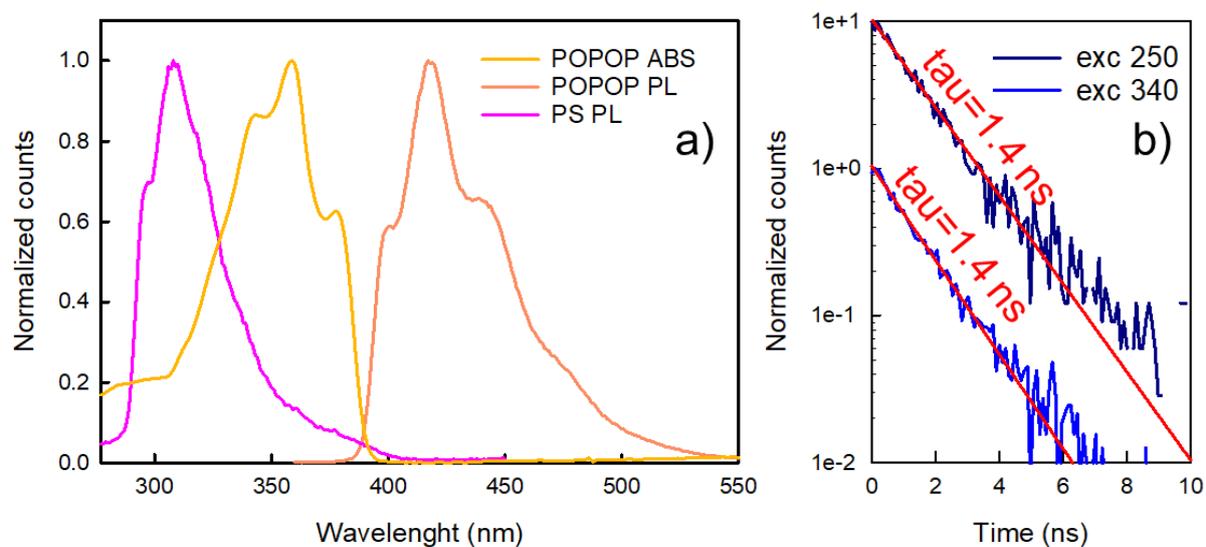

**Figure S1.** a) Absorption and emission spectra of POPOP (yellow line and orange line, respectively) and emission spectra of PS (pink line). The overlap between the matrix emission and the dye absorption is appreciable. b) Time resolved PL signal of POPOP recorded at 420 nm in THF under pulsed excitation at excited at 340 nm (blue) and 250 nm (dark blue). Solid lined are the fit of data with a single exponential decay function.

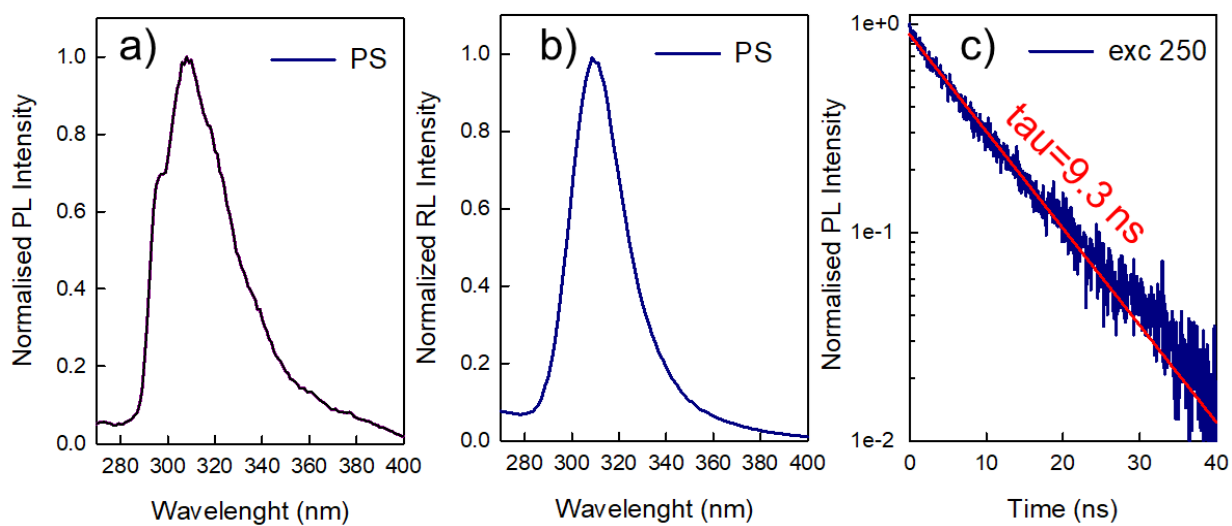

**Figure. S2**. a) Photoluminescence (PL, exc. 250 nm), b) radioluminescence (RL, soft x-rays) and c) 310 nm time resolved PL of monolithic polystyrene.

## 3. PS-to-POPOP resonance energy transfer.

Figure S5a shows the absorption and emission spectra of POPOP dye in THF solution. POPOP has been chosen as acceptor in the Forster energy transfer mechanism due to the overlap of its absorption profile with the polystyrene emission spectra. Figure S6a shows the PL spectra of PS:POPOP sample excited at 250 nm. Here the photoluminescence signal in the UV region of the spectrum ascribed to PS fluorescence is completely missing thus suggesting the presence of a very efficient resonance energy transfer with a $10^{-2}$ M POPOP concentration. The same can be observed with in the radioluminescence spectra reported in Fig.S6b. Figure S6c is reported the photoluminescence lifetime recorded at 310 nm of pure polystyrene (red) and the one of the PS:POPOP sample (black), both excited at 250 nm. The two decays have been measured on sample with the same size and same geometry. The emission intensity has been corrected by the integration time and by the instrumental optical response. From the integrals of the two curves it is possible to calculate the energy transfer efficiency form PS to POPOP as $\phi'_{ET} = 1 - I_{PS:POPOP}/I_{PS} = 0.98$. From the lifetime of the residual PS emission we can estimate the efficiency of the other PS excitons as $\phi''_{ET} = 1 - \tau_{PS:POPOP}/\tau_{PS} = 0.82$.

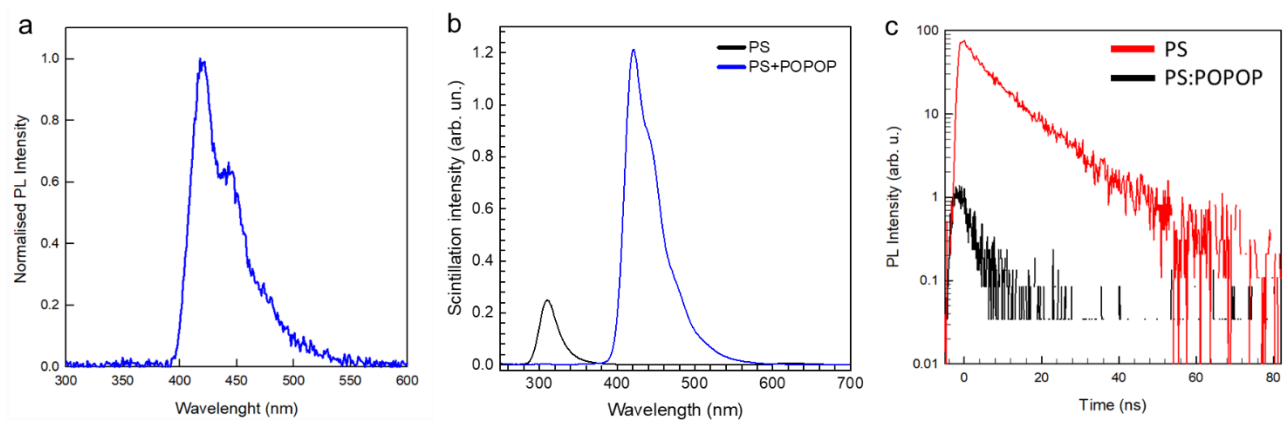

**Figure S3.** a) PL emission of PS:POPOP scintillator excited at 250 nm. The plot shows the absence of the signal related to the PS emission at 310 nm. b) RL spectra of pure PS and PS:POPOP sample. c) PS and PS:POPOP emission intensity decay in time under pulsed excitation at 250 nm.

## 4. Structural Characterization of hafnium oxide nanoparticles.

*Diffraction experiment (XRD).* Powder XRD patterns were acquired in Bragg−Brentano geometry with Cu Kα radiation (analytical X'Pert Pro powder diffractometer).

*RAMAN specroscopy.* Raman spectra was collected using a Labram Dilor spectrometer (JobinYvon) by three accumulations of 100 s of integration. The beam was focused on a circular spot through the optics of a microscope (BX40 Olympus).

*Transmission Electronic Microscopy (TEM) analysis.* Samples were prepared by dispersing a few milligrams of the compounds in 2 mL of deionized water and dropping 3 μL of the solution on carbon-coated copper grids. The samples were analyzed by a JEOL JEM1220 transmission electron microscope operated at 120 kV.

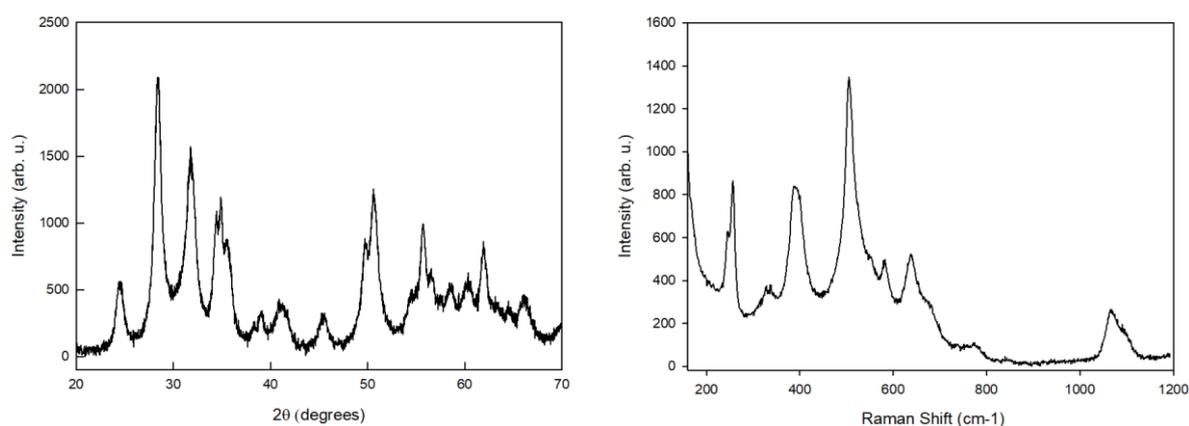

**Figure S4.** a) XRD pattern and b) Raman spectrum of $HfO_2$ nanoparticles employed to fabricate the scintillating nanocomposite.

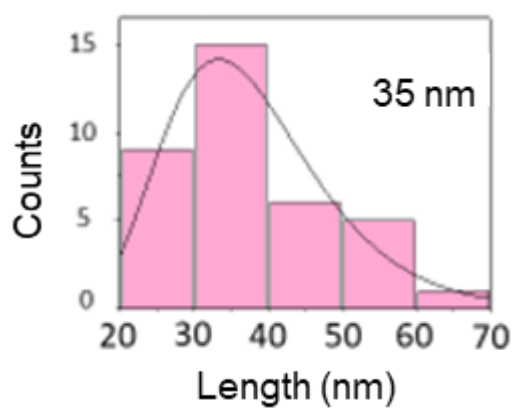

**Figure S5.** Size distribution of the short axis of $HfO_2$ oval nanoparticles obtained by the TEM images analysis.

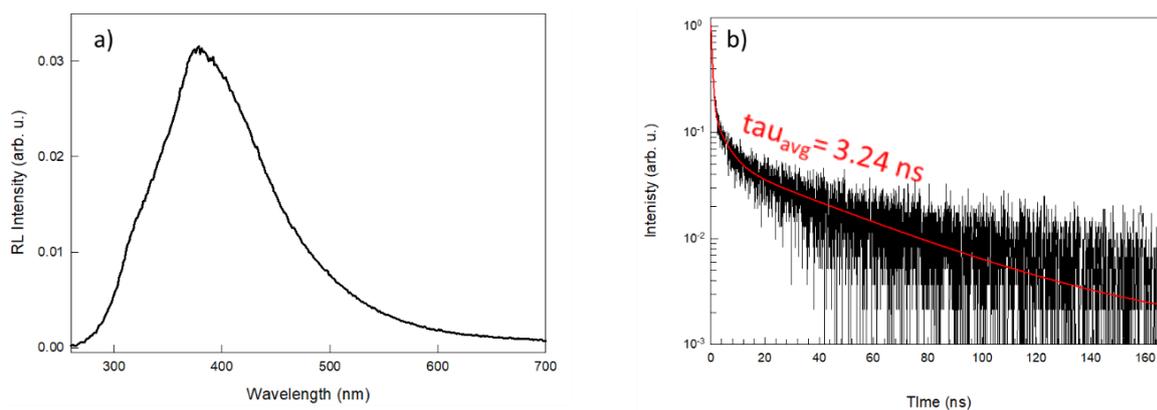

**Figure S6.** a) RL spectrum of HfO$_2$ nanoparticles and b) scintillation under pulsed soft x-ray excitation. The nanoparticle light yield is estimated as low as 100 ph MeV$^{-1}$. The solid line is the fit of data with a multi-exponential decay curve.

**Table. S1.** Fit parameters used to reproduce the hafnium oxide nanoparticles scintillation pulse.

| A$_1$ | τ$_1$ (ns) | A$_2$ | τ$_2$ (ns) | A$_3$ | τ$_3$ (ns) | τ$_{avg}$ (ns) |
|---|---|---|---|---|---|---|
| 0.12 | 4.46 | 0.83 | 0.56 | 0.05 | 42.75 | 3.24 |

## 5. Scintillation of PS:POPOP-2.5% nanocomposites.

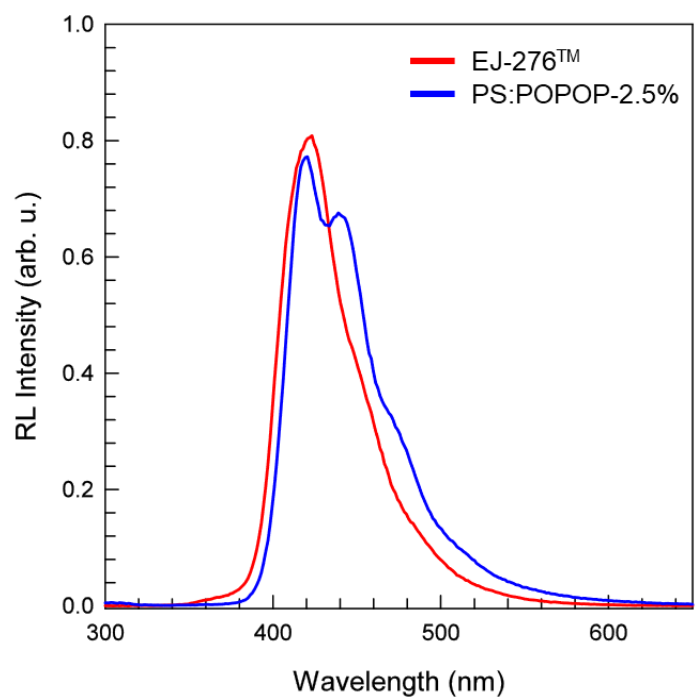

**Figure S7.** Comparison of scintillation emission of EJ-276$^{TM}$ (red) and polystyrene loaded with POPOP ($10^{-2}$ M) and HfO$_2$ nanoparticles (2.5% wt, blue).

**Table S2.** Fit parameters used to reproduced the scintillation pulse intensity decay kinetics under soft x-rays in BGO and PS:POPOP-2.5% 100 mm layers, as well as in the single heterostructures obtained by coupling the two materials as described in the main text.

|  | A$_1$ | $\tau_1$ (ns) | A$_2$ | $\tau_2$ (ns) | $\tau_{avg}$ (ns) |
|---|---|---|---|---|---|
| PS:POPOP-2.5% | 0.7596 | 2.95 | 0.138 | 10.83 | 4.16 |
| BGO | 0.3500 | 42.71 | 0.65 | 321.89 | 224.18 |
| H1 single heteros. | 0.3752 | 36.34 | 0.63 | 310.78 | 210.82 |
| H2 single heteros. | 0.6848 | 4.89 | 0.19 | 244.60 | 57.24 |

## 6. Structural properties PS:POPOP-2.5% nanocomposite.

*SOLID STATE NMR.* $^{13}$C solid-state NMR experiments were carried out at 75.5 MHz with a Bruker NEO 300 instrument operating at a static field of 7.04 T equipped with a 4 mm double resonance MAS probe. $^{13}$C [$^{1}$H] ramped-amplitude Cross Polarization (CP) experiments 5 were performed at 293 K at a spinning speed of 12.5 kHz using a recycle delay of 5 s and contact times of 2 and 0.05 ms. The 90° pulse for proton was 2.9 µs. Crystalline polyethylene was taken as an external reference at 32.8 ppm from TMS.

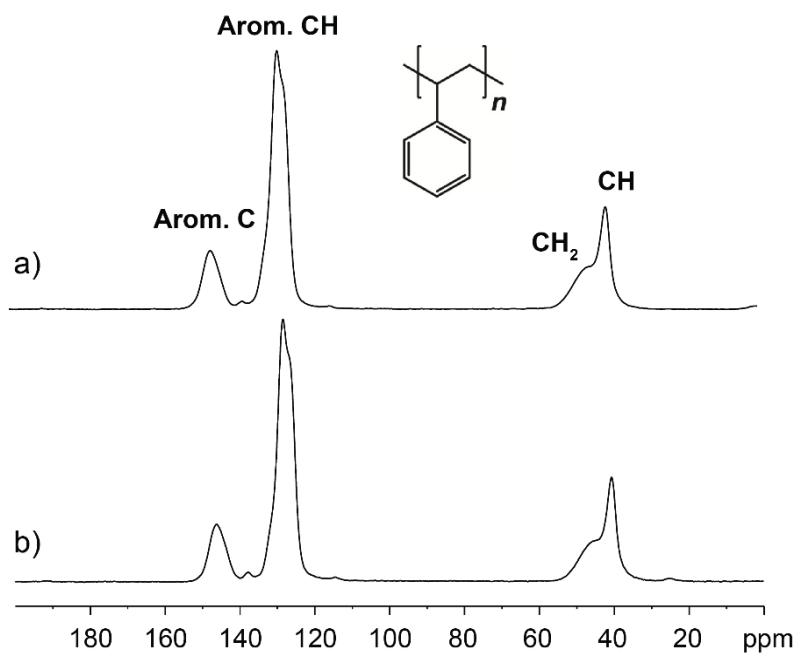

**Figure S8.** $^{13}$C{$^{1}$H} CP spectra of PS-POPOP-2.5% (a) and PS-POPOP (b). The experiments were performed at 293 K at a spinning speed of 11 kHz and 12.5 kHz, respectively, with a contact time of 2 ms.

*DIFFERENTIAL SCANNING CALORIMETRY.* DSC data were recorded on a Mettler Toledo Stare DSC1 analysis system equipped with low temperature apparatus. The experiments were run under nitrogen atmosphere in standard 40 μL Al pans. DSC measurement was performed between 25°C and 150°C at 20°C min$^{-1}$ under nitrogen atmosphere.

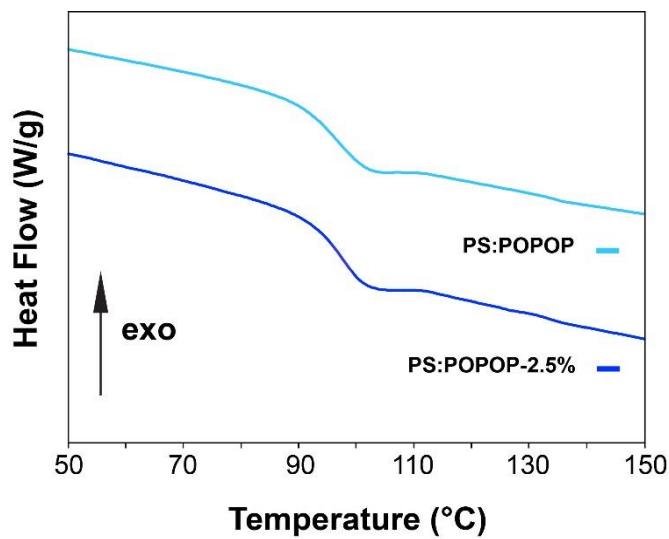

**Figure S9.** DSC thermograms of PS-POPOP-2.5% (blue) and PS-POPOP (light blue). The detection of glass transition in the DSC runs at about 97 °C for PS-POPOP-2.5% and 96 °C for PS-POPOP nanocomposites, respectively, suggested that the same degree of polymerization has been reached for both polymeric materials.

*THERMOGRAVIMETRIC ANALYSIS (TGA)*. TGA were performed using a Mettler Toledo Star System 1 equipped with a gas controller GC10. The experiments were conducted in 70 µL alumina pan applying a thermal ramp from 30°C to 1000°C and a scan rate of 10 °C /min in dry air (50 mL/min).

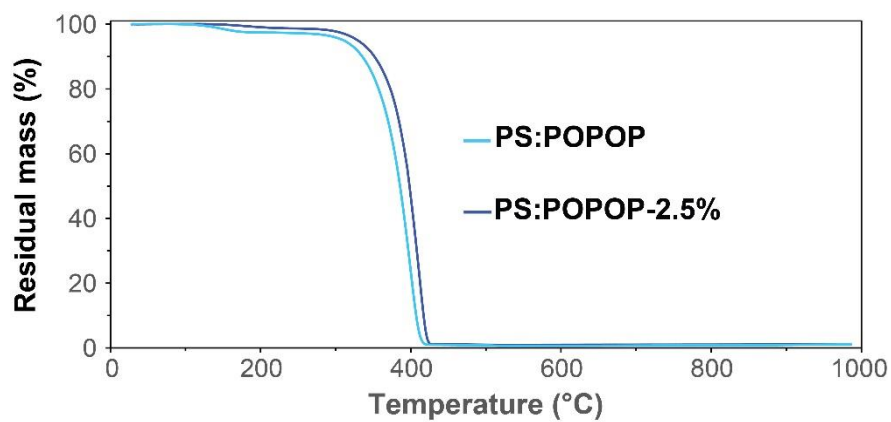

**Figure S10.** Thermogravimetric analysis of PS-POPOP-2.5% (blue) and PS-POPOP (light blue) measured between 30°C and 1000°C under oxidative atmosphere (dry air, 50 mL/min). The weight loss below 200°C is related to removal of small amount of styrene monomer, while the polystyrene-POPOV degradation process starts at about 325°C for PS-POPOP sample and at about 350°C for PS-POPOP-2.5% sample.

# 7. Monte Carlo simulation of the radiation matter interaction in multicomponent scintillators.

A Monte Carlo simulation with the FLUKA code has been developed in order to compute the expected CTR, evaluating the deposited energy within the two heterostructure components as well as the fraction of events depositing an average energy.

Three different setup of the heterostructure have been simulated to study the detector performances. A total detector size of $3\times3\times3$ mm$^3$ has been obtained with:

1) a multilayer where 15 layers of 100 μm thick BGO crystal plates are alternated to 15 layers of 100 μm thick PS:POPOP-2.5% (Multilayer PS:POPOP-2.5% in the following, Fig.S11 left);

2) a multilayer where 15 layers of 100μm thick BGO crystal plates are alternated to 15 layers of 100 μm thick PS (Multilayer PS in the following, Fig.S11a, left);

3) a BGO crystal bulk "drilled" and filled with fiber-type PS:POPOP-2.5% with a radius of 200 μm, height along the detector axis (z axis) of 3 mm (Fiber-type PS:POPOP-2.5% in the following, Fig.S11a, right).

The PS:POPOP-2.5% has a density $\rho = 1.5$ g/cm$^3$. It is composed by a mass fraction of 97.136% PS, 2.5% HfO2 and 0.364% POPOP. The PS ($C_8H_8$) has $\rho = 1.06$ g/cm$^3$, the POPOP ($C_{24}H_{16}N_2O_2$) has $\rho = 1.02$ g/cm$^3$ and the HfO$_2$ has $\rho = 9.68$ g/cm$^3$.

The simulated source is an isotropic gamma source of 511 keV, placed in the origin of the XY reference frame, at a distance from the detector surface of 0.5 cm. The total number of primaries simulated is $1 \times 10^9$, and $\sim 2.6 \times 10^7$, i.e. the $\sim 3$ %, are entering the detector, according to the detector solid angle aperture.

Thanks to a dedicated output developed for the scope, the energy loss in the detector materials has been obtained for the three different setups. Fig. S11a shows the total energy loss ($E_{loss\_tot}$) for the three detector configurations. The fraction of events with a total energy loss $E_{loss\_tot} > 440$ keV has been calculated as the integral of the $E_{loss\_tot} > 440$ keV over the integral of $E_{loss\_tot}$ and it is reported in the last column of the supporting Table S3.

Table S3 reports also the average fraction of energy loss occurring in the BGO and in the polymer for the three considered setup. Such fraction of energy has been calculated event by event as the energy loss in the BGO ($E_{loss\_BGO}$) or the energy loss in the Polymer ($E_{loss\_Poly}$) over $E_{loss\_tot}$, in the case of "shared events", i.e. when $E_{loss\_tot} > 0$ keV, $E_{loss\_BGO} \neq 0$ keV and $E_{loss\_Poly} \neq 0$ keV. The distributions of the fraction of energy deposited in BGO and Polymer are shown in Fig. S11c.

The expected CTR has been evaluated in each deposited energy interval ($E_i$) in the nanocomposite for shared events after the photoelectric events selection (p.e. selection) asking for Eloss_Poly > 50 keV and Eloss_tot > 440 keV. The CTR$_i$ can be inferred from the fraction of shared events in p.e. selection, depositing energy in the E$_i$ interval, with respect to the total number of shared events in p.e. selection.

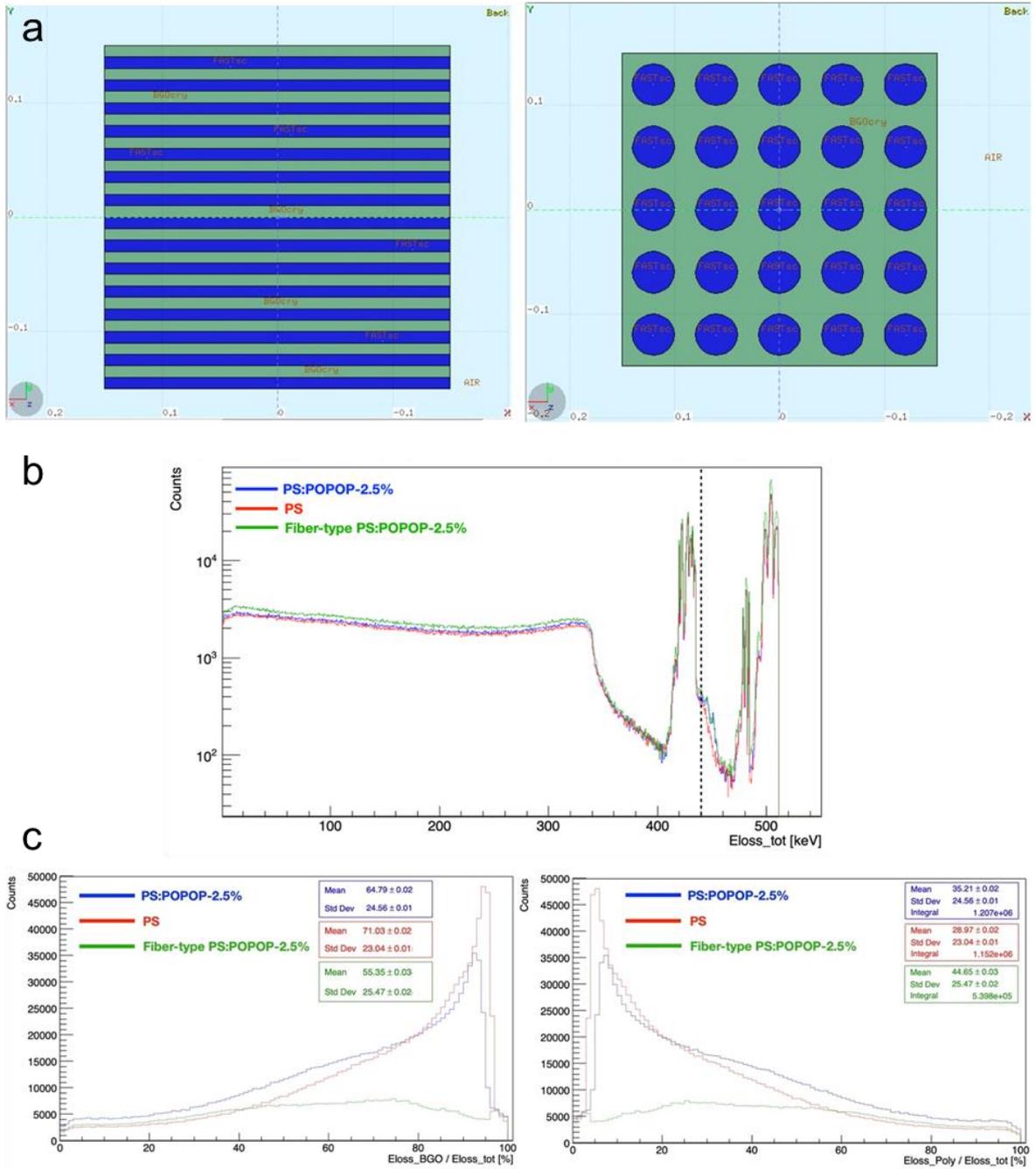

**Fig. S11.** a) Geometry of the two simulated setup in the XY view: the multilayer, where 100 μm thick BGO layers are alternated to 100 μm thick fast scintillator layers (left) and the BGO crystal "drilled" with 200 μm radius, Hz = 3 mm, of fast scintillator fiber-type (right). BGO material is indicated by the green areas, while the fast scintillator material by the blue areas. The total detector size is 3 x 3 x 3 mm$^3$. b) Total energy loss for the three detector configurations: Multilayer PS:POPOP-2.5% (blue line), Multilayer PS (red line) and Fiber-type PS:POPOP-2.5% (green line). The vertical dashed line indicates the $E_{loss\_tot}$ = 440 keV. c) Fraction of Energy in BGO (left) and Polymer (right) for Multilayer PS:POPOP-2.5% (blue line), Multilayer PS (red line) and Fiber-type PS:POPOP-2.5% (green line). The mean of the distributions is shown in the corresponding statistic panel.

**Table S3.** Calculated average fraction of energy occurring in the two components of the scintillators considered taking an isotropic source of 511 keV γ-rays. The Fiber-type geometry employed results a larger relative fraction of polymer with respect tot BGO, so more energy is released in the fast part (ca. +30% with respect to the multiplayer PS:POPOP-2.5%). However, because of the reduced average density and stopping power, the fraction of large energy events useful of imaging reconstruction is basically unchanged, thus making this geometry less effective for ToF-PET applications (see Fig. S12).

|  | **Fraction of Energy in BGO** | **Fraction of Energy in Polymer** | **Fraction of Events with energy loss > 440 keV** |
|---|---|---|---|
| Multilayer PS:POPOP-2.5% | 0.65 | 0.35 | 0.24 of the total |
| Multilayer PS | 0.71 | 0.29 | 0.24 of the total |
| Fiber-type PS:POPOP-2.5% | 0.55 | 0.45 | 0.29 of the total |

**Table S4.** Calculated cumulative probability $A_i$ distribution of shared events in photoelectric event selection as a function of the fraction of energy $E$ deposited in the nanocomposite $\varepsilon_i$ (relative to the maximum 511 keV) considering an isotropic source of 511 keV γ-rays.

|  |  | **Multilayer PS:POPOP-2.5%** | **Multilayer PS** |  | **Multilayer PS:POPOP-2.5%** |
|---|---|---|---|---|---|
| $E_i$ deposited in the nanocomposite | $\varepsilon_i$ ($E_i$/ 511 $keV$) | $A_i$ | $A_i$ |  | $CTR_i$ (ps) |
| 50 - 150 keV (avg. 100 keV) | 0.19 | 0.52 | 0.62 |  | 104 |
| 150 - 250 keV (avg. 200 keV) | 0.39 | 0.33 | 0.29 |  | 84 |
| 250 - 350 keV (avg. 300 keV) | 0.59 | 0.13 | 0.07 |  | 78 |
| 350 - 450 keV (avg. 400 keV) | 0.78 | 0.02 | 0.01 |  | 78 |
| 450 - 550 keV (avg. 500 keV) | 0.98 | 0.001 | 0.0003 |  | 85 |
| Events with significant fast emission ($E \geq 150$ keV) |  | 48% | 38% | TOTAL $CTR^†$ | 94 |

## 8. Additional Data.

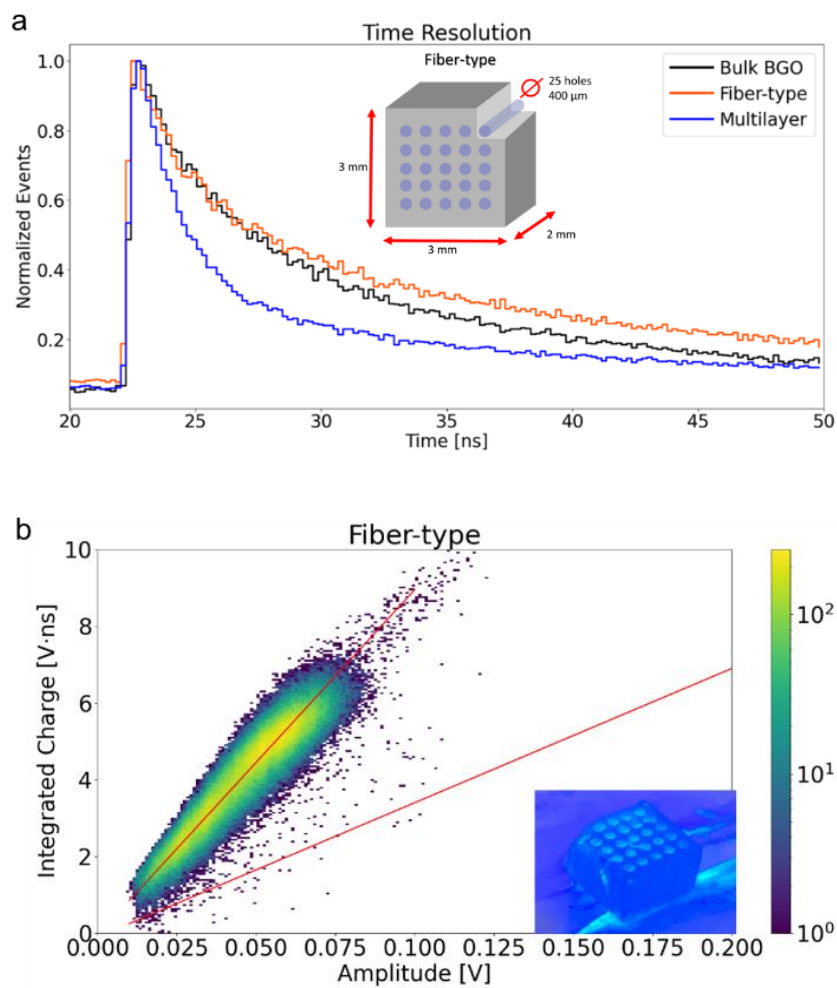

**Figure S12.** a) Detector time resolution and b) time integrated pulse-height spectrum of 3x3x3 mm³ fiber-type scintillator obtained by filling a drilled BGO cube with the PS:POPOP-2.5% nanocomposite. The inset show a sketch of the scintillator structure and a digital picture of it under UV lamp exposure, in panel a and b, respectively. We cannot observe the intermediate response in time as like in the multilayer case, nor by exciting with soft- rays, nor by using 511 keV γ-rays.

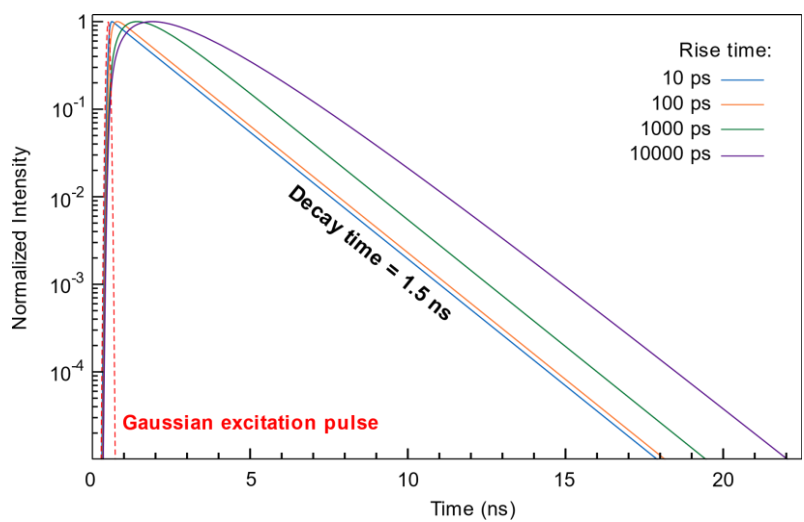

**Figure S13.** Simulated light pulse generated form POPOP calculated as a function of the rise time, from 10 to 10000 ps, and considering a convoluted Gaussian shape excitation pulse (dashed line) with a FWHM of 110 ps. The emission rise time has been modulated in order to simulate a PS-to-POPOP energy transfer of different rate that activates the POPOP emission upon ultrafast laser excitation. The emission decay time does not change, the slow energy transfer only delays the time at which the emission reaches its maximum.